\documentclass[twocolumn,showpacs,preprintnumbers,amsmath,amssymb]{revtex4}

\usepackage{graphicx}
\usepackage{dcolumn}
\usepackage{bm}

\begin{document}

\preprint{APS/123-QED}

\title{Medium mass fragments production due to momentum dependent interactions\\}

\author{Sanjeev Kumar}
\author{Suneel Kumar}%
 \email{suneel.kumar@thapar.edu}
\affiliation{%
School of Physics and Material Science, Thapar University Patiala-147004, Punjab (India)\\
}%

\author{Rajeev K. Puri}
\affiliation{
Department of Physics, Panjab University, Chandigarh (India)\\
}%

\date{\today}

\begin{abstract}
The role of system size and momentum dependent effects are analyzed in multifragmenation
by simulating symmetric reactions of Ca+Ca, Ni+Ni, Nb+Nb, Xe+Xe, Er+Er, Au+Au, and U+U at incident
energies between 50 MeV/nucleon and 1000 MeV/nucleon and over full impact parameter zones.
Our detailed study reveals that there exist a system size dependence when reaction is simulated with
momentum dependent interactions. This dependence exhibits a mass power law behavior.\\
\end{abstract}

\pacs{25.70.Pq, 25.70.-z, 24.10.Lx}
\keywords{momentum dependent interactions, quantum molecular dynamics, medium mass fragments, multifragmentation}
\maketitle

\section{Introduction}
The last gem in the field of heavy-ion collisions, namely, multifragmentation has always attracted theoreticians as well as
experimentalists \cite{Aich91,Ogil91,Li93,Maru97,Li06}. Primarily due to the several hidden phenomena that need deeper investigations and secondary due to
its connection with nuclear equation of state - {\it a question which always has captured a central place in the research
of nuclear physics}. The knowledge of the nuclear compressibility is not only relevant for the nuclear physics, it is also
vital for other branches such as astrophysics. One should, however, note that the compressibility depends not
only on the density but also the entire momentum plane. In other words, equation of state apart from the population
of nucleons also depends upon their relative velocities. This can also been seen from the optical potential where
strong momentum dependence was reported in the literature\cite{Aich87}.\\
The momentum dependence of the nuclear equation of state has
been reported to affect the collective flow and particle production drastically \cite{Aich91,Huan93,Kuma99a,Sing01,Sood06,
Sing02, Zuo05,Li04,Chen07,Dani00}. Some initial investigations also point
toward its important role in multifragmentation \cite{Kuma99a,Sing01,Colo07,Ono04}. Due to the repulsive nature of
momentum dependent interactions (MDI), nuclei
propagating with MDI are reported to emit nucleons and light complex fragments.
However, the rms radii of single nuclei is not affected significantly \cite{Kuma99a}.
One has to keep in the mind that the response of momentum dependent interactions also
depends on the system size. For example, it has been shown by Sood and Puri \cite{Sood06} that momentum dependent interactions push
energies of vanishing flow to significant lower levels for $^{12}C + ^{12}C$ system, whereas
for heavier systems, the trend is just opposite.
In a recent study, Singh and Puri \cite{Sing02} predicted a power law for the
system size effects in multifragmentation. In their study, a simple static equation of state was used.
In view of the above facts, it is challenging to investigate the role of momentum dependent interactions with respect
to multifragmentation $\&$ see how system size affects the outcome. Our present aim, therefore, is three fold atleast:\\
(i) to understand the role of momentum dependent interactions in multiframentation,\\
(ii) to study the system size effects in the presence of momentum dependent interactions and
(iii) to find a scaling to these system size effects. \\
This study is done within the frame work of quantum molecular dynamics model. The section II deals with the model, section III
discuss the results. Our results are summarized in section IV.\\
\section{The model}
The QMD model is a time dependent many-body theoretical approach which is
based on the molecular dynamics picture that treats nuclear correlations explicitly.
The two dynamical ingredients of the model are the density dependent mean field and the
in-medium nucleon-nucleon cross-section
\cite{Bohn89}.
In the QMD model, each nucleon is represented by a Gaussian wave packet characterized by the
time dependent parameters in space $\vec{r_i}(t)$ and in momentum $\vec{p_i}(t)$\cite{Aich91}.
This wave packet can be represented as:
\begin{equation}
\Phi_{i}({\vec{r},\vec{p},t})~=~ \frac{1}{(2\pi L)^{3/4}}\exp^{-(\vec{r}-\vec{r_i}(t))^{2}/2L}
\exp^{\iota \vec{p_i}(t).\vec{r}/\hbar}~~\cdot
\end{equation}
The total n-body wave function is assumed to be a direct product of the form:
\begin{equation}
\Psi~=~\Pi_{i=1}^{A_{T}+A_{P}}\Phi_{i}~~\cdot
\end{equation}
The model uses its classical analog in term of Wigner function \cite{Carr83}.
\begin{equation}
f_{i}({\vec{r},\vec{p},t})~=~\frac{1}{(\pi\hbar)^3}\exp^{-[\vec{r}-\vec{r_i}(t)]^{2}/4L}
\cdot\exp^{-[\vec{p}-\vec{p_i}(t)]^2\cdot{2L}/\hbar^2}~~\cdot
\end{equation}
The parameter L is related to the extension of the wave packet in phase space. This parameter is, however,
kept fixed in the present study.
The centroid of each nucleon propagates under the classical equations of motion \cite{Aich91}:
\begin{equation}
\frac{d\vec{r_i}}{dt}~=~\frac{d\it{H}}{d\vec{p_i}}~~;~~\frac{d\vec{p_i}}{dt}~=~-\frac{d\it{H}}
{d\vec{r_i}}~~\cdot
\end{equation}
The $H$ referring to the Hamiltonian reads as:
\begin{equation}
\it{H}~=~\sum_{i}\frac{\vec{p_i}^2}{2m_i}~+~V^{tot}~~\cdot
\end{equation}
Our total interaction potential $V^{tot}$ is a composite of various terms:
\begin{equation}
V^{tot}~=~V^{loc}+V^{Yukawa}+V^{Coul}+V^{MDI}~~\cdot
\label{s1}
\end{equation}
with
\begin{equation}
V^{loc}~=~t_{1}\delta({\vec{r_1}-\vec{r_2}})+t_{2}\delta(\vec{r_1}-\vec{r_2})\delta(\vec{r_1}-\vec{r_3}),
\end{equation}
\begin{equation}
V^{Yukawa}~=~t_{3}\frac{exp(-|\vec{r_1}-\vec{r_2}|/m)}{|\vec{r_1}-\vec{r_2}|/m},
\end{equation}
and
\begin{equation}
V^{MDI}~\approx~t_4ln^2[t_5(\vec{p_1}-\vec{p_2})^2+1]\delta(\vec{r_1}-\vec{r_2})~~\cdot
\end{equation}
Here $m = 1.5$ fm, $t_3 = -6.66$ MeV, $t_4 = 1.57$ MeV and $t_5 = 5\times10^{-4}$ MeV$^{-2}$.\\
The momentum dependent interactions can be incorporated by parameterizing
the momentum dependence of the real part of optical potential \cite{Aich87}.\\
\section{Results and Discussion}
\label{sec:2}
We simulated each reaction at various time steps and stored the phase space. This phase space needs to be
clusterized. In the present study, we clusterized the phase space using the
minimum spanning tree (MST) method which binds two nucleons in a fragment if their
centroids are closer than 4 fm. In other words, we demand:
\begin{equation}
|\vec{r_i}-\vec{r_j}|~\leq~4~~ fm~~\cdot
\label{s3}
\end{equation}
At present, various reactions were simulated with soft and
soft momentum dependent (MDI) equations of state (EOS) with comressibility $K = 200 MeV$. 
A standard energy dependent nucleon -nucleon cross-section due
to Cugnon was also used
\cite{Aich91}.
 In brief, we followed the time evolution till the end of the reaction which, in the present study, is
200 fm/c. \\
Here each of the reaction $^{40}_{20}Ca+^{40}_{20}Ca$,
$^{58}_{28}Ni+^{58}_{28}Ni$, $^{93}_{41}Nb+^{93}_{41}Nb$,
$^{131}_{54}Xe+^{131}_{54}Xe$, $^{167}_{68}Er+^{167}_{68}Er$, $^{197}_{79}Au+^{197}_{79}Au$ and
$^{238}_{92}U+^{238}_{92}U$ was simulated for 100 events at incident energies
between 50 and 1000 MeV/nucleon using different collision geometries.
 By using the symmetric (colliding) nuclei, system size effects can be analyzed without varying the
 asymmetry (and excitation energy) of the system. It is worth mentioning that the experimental studies by the MSU mini ball
and ALADIN \cite{Ogil91,Bowm92} groups, varied the asymmetry of the reaction whereas plastic ball
 \cite{Gutb90} and FOPI experiments \cite{Scha97}
are performed for the symmetric colliding nuclei only. In the following, we first discuss the time evolution of different
reactions
and then shall address the question of momentum dependent interactions and system size effects.\\
\subsection{Time evolution}
\begin{figure}
\includegraphics{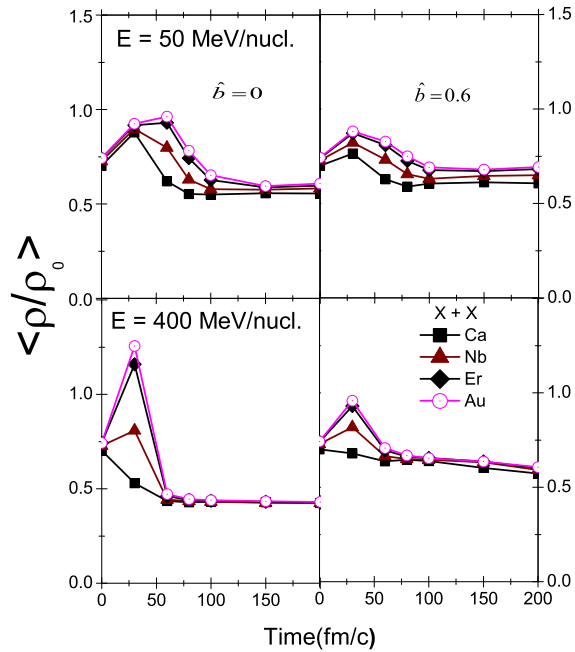}
\caption{\label{fig:1} (Color online) Average density $ \langle{\rho/\rho_{0} \rangle}$ as a function of the time.
The top panel is at 50 MeV/nucleon, where bottom panel represents the reaction at 400 MeV/nucleon. The left and
right hand sides represent, respectively, the central collision $\hat{b}~=~0$ and peripheral collision
$\hat{b}~=~0.6$. All reactions represent symmetric colliding nuclei $X~+~X$, where X represents the reacting elements.}
\end{figure}
The density of the environment surrounding the nucleons of a fragment plays crucial role in deciding the physical process
behind their formation. In fig.\ref{fig:1}, we display the average 
density $<\rho/\rho_{0}>$ reached in a typical reaction as a function of the time.
The average nucleonic density $<\rho/\rho_{0}>$ is calculated as \cite{Bohn89}
\begin{eqnarray}
<\rho/\rho_{0}>~=~\langle\frac{1}{A_T~+~A_P}& &\sum_{i=1}^{A_T~+~A_P}\sum_{j=1}^{A_T~+~A_P}
\frac{1}{(2\pi L)^{3/2}}\nonumber\\
& &\cdot exp[-(\vec{r_i}-\vec{r_j})^{2}/2L]\rangle,
\end{eqnarray}
with $\vec{r_i}$ and  $\vec{r_j}$ being the position coordinates of $i^{th}$ and $j^{th}$ nucleons, respectively.
We here display the average density at incident energies of 50 and 400 MeV/nucleon. In addition,
two colliding geometries corresponding to $\hat{b}~=~0$
and $\hat{b}~=~0.6$ are also taken. The reaction (at low incident energies) preserves most of the initial
correlations and hence only a small change in the density profile occurs. This trend turns to a sharp decrease
at higher incident energies. This is due to the fact that higher incident energies lead to highly unstable
compressed zone which, does not sustain for a long time and as a result fast emission of nucleons occurs.\\
A similar trend is also seen for the collision profile which has a direct relation with the density
reached in a reaction.\\
\begin{figure}
\includegraphics{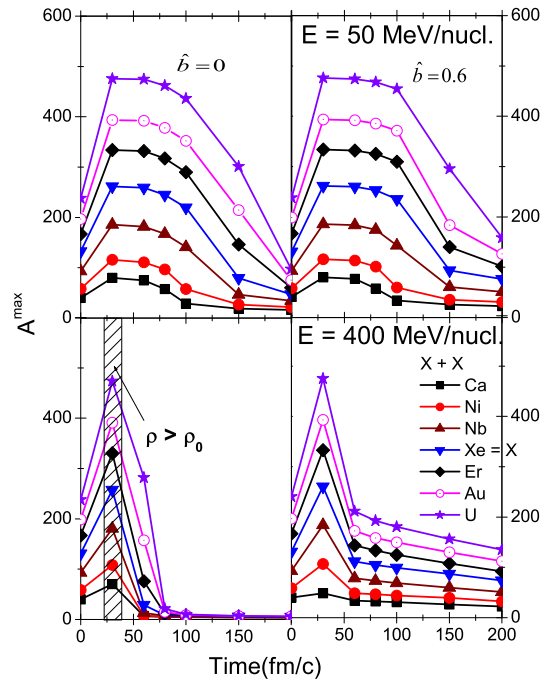}
\caption{\label{fig:2} (Color online) Same as fig.\ref{fig:1}, but for the time evolution of the heaviest fragment $A^{max}$ as a function of the time. The shaded area corresponds to density higher than normal nuclear matter density $\rho_{0}$.}
\end{figure}
\begin{figure}
\includegraphics{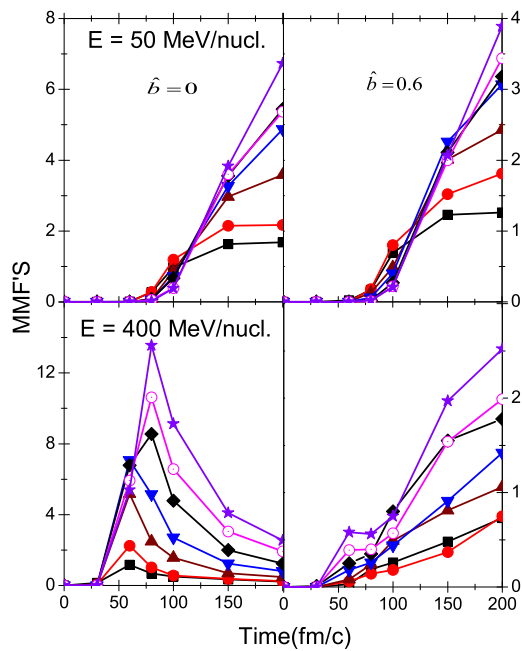}
\caption{\label{fig:3} (Color online) Same as in fig.\ref{fig:1}, but for the  time evolution of multiplicity of MMF's.}
\end{figure}
Let us now turn towards multifragmentation. In fig.\ref{fig:2}, we display the heaviest fragment $A^{max}$
survived in a reaction. The
medium mass fragments (MMF's), defined as 5 $\le$ A $\le$ 9, are displayed in fig.\ref{fig:3}. The top panel
in both figures is for 50 MeV/nucleon, whereas
bottom panel is at 400 MeV/nucleon. The time evolution of the fragments reveals many interesting points:
The heaviest fragment $A^{max}$ survived in the reaction of heavier systems struggles for a longer time.
The soft momentum dependent (SMD) equation of state destroy most of the nucleon-nucleon correlations,
 therefore, $A^{max}$ obtained with SMD EOS is lighter than corresponding soft equation of state \cite{Sing02}. Consequently, there is an enhanced emission of the free nucleons, LMF's (light mass fragments) and MMF's.
We also see an appreciable enhancement in the nucleonic emission (not shown here) at all incident energies and
 impact parameters. For heavier systems, emission of nucleons continues
 till the end of the reaction. This is due to the finite collisions happening at the later stage as well as
due to the longer reaction time at these incident energies.\\
It takes longer time for $A^{max}$ in heavy systems to be stabilized compared to lighter
nuclei, where saturation time for $A^{max}$ is much less. The excited $A^{max}$ in heavier system
continues to emit the nucleons till the end of the reaction. This time is, however, much shorter in lighter nuclei.\\
The multiplicity of MMF's has a different story to tell.
We now see more fragments in central collisions at $50$ MeV/nucleon compared to peripheral collisions.
As we increase the energy, the
MMF's production decreases considerably. This is valid for the central collisions only.
The peripheral collisions yield almost same MMF's. As noted in ref. \cite{Sing02}, the static soft EOS is not able to
break the initial correlations among
the nucleons in peripheral collisions. As soon as momentum dependence is taken into account,
the initial correlations among the nucleons are destroyed,
resulting in large number of MMF's. The simple static EOS
fails to transfer the energy from the participant to spectator matter. In other words, MDI suppresses
the production of free nucleons and LMF's while the production of MMF's is enhanced.
Overall, we observe enhancement in the multiplicity of medium mass fragments with MDI compared
to static equation of state.\\
\subsection{Final state fragment distribution}
As we know, measurements are always done at the end of the reaction. The reaction
time is chosen to be $t = 200$ fm/c. It is based on the fact that directed flow
saturates by this time. Therefore, it will
be of interest to see whether the final state fragment distribution has momentum and system size dependence
or not.
\begin{figure}
\includegraphics{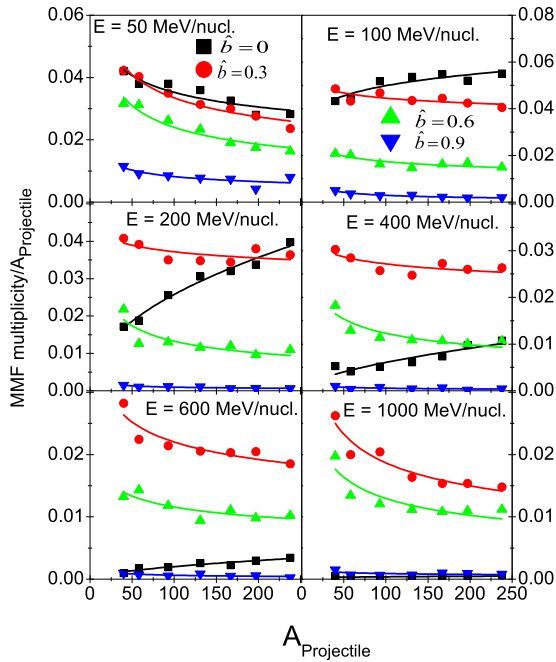}
\caption{\label{fig:4} (Color online) Final state multiplicity of the medium mass fragments per 
projectile nucleon as a function of projectile mass $A_{P}$. 
The top, middle and bottom panel at left side represent, respectively, the reactions at 50, 200, 600 MeV/nucleon.
The corresponding right hand side represent, respectively, the reactions  at
100, 400, 1000 MeV/nucleon . All curves are fitted using function $Y~=~CA^\tau_{P}$.}
\end{figure}
We display in fig.\ref{fig:4}, the reduced multiplicity (multiplicity/nucleon) of medium mass fragments (MMF's).
In experimental observations (e.g.
FOPI and ALADIN), the nuclear matter is divided into spectator and participant parts \cite{Zbir07}. In our calculations,
this division is made by splitting the reaction into different impact parameter zones that can be
related with the spectators/participant matter. The top panel in fig.\ref{fig:4} displays the multiplicity of
MMF's at 50 and 100 MeV/nucleon,
whereas bottom panel is at 600 and 1000 MeV/nucleon, respectively. The middle panel represents the outcomes at
200 and 400 MeV/nucleon.
Each window of the panel contains four different curves that correspond, respectively, to the
scaled impact parameter values of $\hat{b}= 0.0, 0.3, 0.6, 0.9$. The wide range of incident energy between 50 and
1000 MeV/nucleon and impact parameter between zero and $b_{max}$ gives opportunity to study the different dynamics
emerging at low, intermediate and higher energies. \\
\begin{figure}
\includegraphics{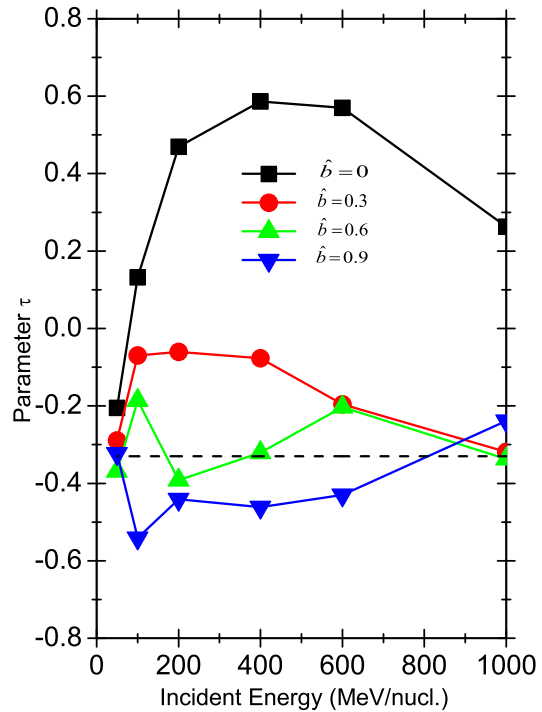}
\caption{\label{fig:5} (Color online) Parameter $\tau$ (appearing in the power law 
function $A_{P}$) as a function of incident energy.
The panel displays the value of $\tau$ for medium mass fragments.}
\end{figure}
The energy received by the target in peripheral collisions is not enough to excite the matter above the Fermi level,
resulting in fewer light fragments. As a result, emission of heavier fragments takes place.
Irrespective of the incident energy and impact parameter, the multiplicity of medium mass fragments
(5 $\le$ A $\le$ 9) scales with the size of
the projectile that can be parametrized by a power law of the form $C(A_{P})^{\tau}$; $A_{P}$ 
is the mass of the projectile. The values of C and $\tau$ depends on the size of fragments as well as on the incident energy
and impact parameter of the reaction.\\
In fig.\ref{fig:5}, we plot the power law  parameter $\tau$ as a function of incident energy and
impact parameter. Although, no unique dependence occurs for $\tau$, we can correlate some of its values.
No physical correlation can be extracted for the central collisions. This is perhaps due to the complete
destruction of initial correlations, moreover, as a result, even no collective flow has been observed in central collisions \cite{Reis97}.\\

\section{Conclusion}
We have studied the role of momentum dependent interactions in fragmentation by
systematically analyzing various reactions at incident energies between 50  and 1000 MeV/nucleon and over full geometrical
overlap.
The inclusion of momentum dependent interactions leads to less freeze out density in all colliding systems.
This happens due to repulsive nature of momentum dependent interactions.
The system size effects are found to vary with reaction parameters and incident energies.
The multiplicity of medium mass fragments can be parametrized in term of a power law. 
This is true for a wide range of
impact parameters and incident energies considered here. However, the parameter $\tau$ 
does not have unique value.\\
\begin{acknowledgments}
This work has been supported by the Grant no. 03(1062)06/ EMR-II, from the Council of Scientific and
Industrial Research (CSIR) New Delhi, Govt. of India.
\end{acknowledgments}

\end{document}